\documentclass[12pt]{article}
\usepackage{amsfonts}
\usepackage{amssymb}
\usepackage{graphics,psboxit,amsmath}

\def\hybrid{\topmargin -20pt    \oddsidemargin 0pt
        \headheight 0pt \headsep 0pt
        \textwidth 6.35in       
        \textheight 9.25in       
        \marginparwidth .875in
        \parskip 5pt plus 1pt   \jot = 1.5ex}

\hybrid

\def\baselinestretch{1.2}

\catcode`\@=11

\def\marginnote#1{}
\newcount\hour
\newcount\minute
\newtoks\amorpm
\hour=\time\divide\hour by60
\minute=\time{\multiply\hour by60 \global\advance\minute by-\hour}
\edef\standardtime{{\ifnum\hour<12 \global\amorpm={am}%
        \else\global\amorpm={pm}\advance\hour by-12 \fi
        \ifnum\hour=0 \hour=12 \fi
        \number\hour:\ifnum\minute<10 0\fi\number\minute\the\amorpm}}
\edef\militarytime{\number\hour:\ifnum\minute<10 0\fi\number\minute}

\def\draftlabel#1{{\@bsphack\if@filesw {\let\thepage\relax
   \xdef\@gtempa{\write\@auxout{\string
      \newlabel{#1}{{\@currentlabel}{\thepage}}}}}\@gtempa
   \if@nobreak \ifvmode\nobreak\fi\fi\fi\@esphack}
        \gdef\@eqnlabel{#1}}
\def\@eqnlabel{}
\def\@vacuum{}
\def\draftmarginnote#1{\marginpar{\raggedright\scriptsize\tt#1}}

\def\draft{\oddsidemargin -.5truein
        \def\@oddfoot{\sl preliminary draft \hfil
        \rm\thepage\hfil\sl\today\quad\militarytime}
        \let\@evenfoot\@oddfoot \overfullrule 3pt
        \let\label=\draftlabel
        \let\marginnote=\draftmarginnote
   \def\@eqnnum{(\theequation)\rlap{\kern\marginparsep\tt\@eqnlabel}%
\global\let\@eqnlabel\@vacuum}  }

\def\preprint{\twocolumn\sloppy\flushbottom\parindent 2em
        \leftmargini 2em\leftmarginv .5em\leftmarginvi .5em
        \oddsidemargin -.5in    \evensidemargin -.5in
        \columnsep .4in \footheight 0pt
        \textwidth 10.in        \topmargin  -.4in
        \headheight 12pt \topskip .4in
        \textheight 6.9in \footskip 0pt
        \def\@oddhead{\thepage\hfil\addtocounter{page}{1}\thepage}
        \let\@evenhead\@oddhead \def\@oddfoot{} \def\@evenfoot{} }

\def\numberbysection{\@addtoreset{equation}{section}
        \def\theequation{\thesection.\arabic{equation}}}

\def\underline#1{\relax\ifmmode\@@underline#1\else
        $\@@underline{\hbox{#1}}$\relax\fi}

\def\titlepage{\@restonecolfalse\if@twocolumn\@restonecoltrue\onecolumn
     \else \newpage \fi \thispagestyle{empty}\c@page\z@
        \def\thefootnote{\fnsymbol{footnote}} }

\def\endtitlepage{\if@restonecol\twocolumn \else \newpage \fi
        \def\thefootnote{\arabic{footnote}}
        \setcounter{footnote}{0}}  

\catcode`@=12
\relax

\def\figcap{\section*{Figure Captions\markboth
        {FIGURECAPTIONS}{FIGURECAPTIONS}}\list
        {Figure \arabic{enumi}:\hfill}{\settowidth\labelwidth{Figure
999:}
        \leftmargin\labelwidth
        \advance\leftmargin\labelsep\usecounter{enumi}}}
 \relax
\def\tablecap{\section*{Table Captions\markboth
        {TABLECAPTIONS}{TABLECAPTIONS}}\list
        {Table \arabic{enumi}:\hfill}{\settowidth\labelwidth{Table
999:}
        \leftmargin\labelwidth
        \advance\leftmargin\labelsep\usecounter{enumi}}}
 \relax
\def\reflist{\section*{References\markboth
        {REFLIST}{REFLIST}}\list
        {[\arabic{enumi}]\hfill}{\settowidth\labelwidth{[999]}
        \leftmargin\labelwidth
        \advance\leftmargin\labelsep\usecounter{enumi}}}
 \relax
\makeatletter
\newcounter{pubctr}
\def\publist{\@ifnextchar[{\@publist}{\@@publist}}
\def\@publist[#1]{\list
        {[\arabic{pubctr}]\hfill}{\settowidth\labelwidth{[999]}
        \leftmargin\labelwidth
        \advance\leftmargin\labelsep
        \@nmbrlisttrue\def\@listctr{pubctr}
        \setcounter{pubctr}{#1}\addtocounter{pubctr}{-1}}}
\def\@@publist{\list
        {[\arabic{pubctr}]\hfill}{\settowidth\labelwidth{[999]}
        \leftmargin\labelwidth
        \advance\leftmargin\labelsep
        \@nmbrlisttrue\def\@listctr{pubctr}}}
 \relax
\makeatother
\newskip\humongous \humongous=0pt plus 1000pt minus 1000pt

\newif\ifdtup

\relax

\def\be{\begin{equation}}
\def\ee{\end{equation}}
\def\ba{\begin{eqnarray}}
\def\ea{\end{eqnarray}}

\def\a{\alpha}

\def\l{\lambda}

\def\no{\noindent}

\def\IR{\relax{\rm I\kern-.18em R}}
\def\II{\relax{\rm 1\kern-.35em1}}

\newcommand{\xvar}{x}

\newcommand{\xp}[1]{\xvar^+_{#1}}
\newcommand{\xm}[1]{\xvar^-_{#1}}
\newcommand{\xpm}[1]{\xvar^\pm_{#1}}

\newcommand{\bernoulli}{\mathrm{B}}
\renewcommand{\digamma}{\mathop{\smash{\oldPsi}\vphantom{a}}\nolimits}
\newcommand{\sfrac}[2]{{\textstyle\frac{#1}{#2}}}
\newcommand{\half}{\sfrac{1}{2}}

\def\IR{\relax{\rm I\kern-.18em R}}
\def\inv{^{\raise.15ex\hbox{${\scriptscriptstyle -}$}\kern-.05em 1}}

\begin{document}

\begin{titlepage}
\begin{center}

\hfill CERN-PH-TH/2006-229\\
\vskip -.1 cm
\hfill IFT-UAM/CSIC-06-57\\
\vskip -.1 cm
\hfill hep--th/0611014\\

\vskip .5in

{\LARGE Integrability and non-perturbative effects in the AdS/CFT correspondence}
\vskip 0.4in

{\bf C\'esar G\'omez}$^{1,2}$\phantom{x}and\phantom{x}{\bf Rafael Hern\'andez}$^{1,2}$
\vskip 0.1in

${}^1\!$
Instituto de F\'{\i}sica Te\'orica UAM/CSIC\\
Universidad Aut\'onoma de Madrid,
Cantoblanco, 28049 Madrid, Spain\\
{\footnotesize{\tt cesar.gomez@uam.es}}
    
\vskip .2in
  
${}^2\!$
Theory Group, Physics Department, CERN\\
CH-1211 Geneva 23, Switzerland\\
{\footnotesize{\tt rafael.hernandez@cern.ch}}

\end{center}

\vskip .4in

\centerline{\bf Abstract}
\vskip .1in
\no
We present a non-perturbative resummation of the asymptotic strong-coupling expansion for 
the dressing phase factor of the $AdS_5 \times S^5$ string S-matrix. The non-perturbative 
resummation provides a general form for the coefficients in the weak-coupling expansion, 
in agreement with crossing symmetry and transcendentality. The ambiguities of the non-perturbative 
prescription are discussed together with the similarities with the non-perturbative definition 
of the $c=1$ matrix model.

\noindent

\vskip .4in
\noindent

\end{titlepage}
\vfill
\eject

\def\baselinestretch{1.2}


\baselineskip 20pt


\no
{\bf Introduction.} The uncovering of integrable structures on both sides of the 
AdS/CFT correspondence \cite{AdS} has suggested a path toward a complete formulation of 
the duality. On the gauge theory side, focussing on operators with large quantum 
numbers \cite{BMN} lead to the identification of the planar dilatation operator of ${\cal N}=4$ 
supersymmetric Yang-Mills with the hamiltonian of an integrable spin chain \cite{int,higherint}. 
Assuming that integrability holds at higher orders a long-range Bethe ansatz was then 
proposed to describe the spectrum of Yang-Mills operators~\cite{longrange}. Classical 
integrability of type IIB string theory on $AdS_5 \times S^5$~\cite{Polchinski} 
allowed a resolution of the sigma model spectrum in terms of spectral curves \cite{Kazakov}, 
and suggested a discrete set of Bethe equations for the quantum string sigma model 
\cite{AFS,completeqBethe}. Integrability on each side of the correspondence is thus encoded 
in an asymptotic factorizable S-matrix satisfying the Yang-Baxter equation. However the 
Yang-Baxter relations do not completely constrain the S-matrix, and it can only be fixed up 
to a scalar dressing phase factor \cite{BeisertS}. The dressing phase factor of the S-matrix 
could be determined by requiring some sort of crossing invariance \cite{Janik}. The structure 
of this dressing phase factor modifies in such a way the long-range Bethe ansatz equations 
that if it remained non-trivial in the weak-coupling regime it would induce perturbative 
violations of the BMN-scaling limit. One interesting feature of the long-range Bethe ansatz 
for high twist operators is that, assuming a trivial dressing factor, it agrees \cite{Eden} 
with the Kotikov-Lipatov transcendentality principle \cite{KL}. Moreover, it is possible to 
have non-trivial dressing phase factors that violate perturbative BMN-scaling but still preserve 
the transcendentality structure \cite{BES}. 
  
The dressing phase factor has been argued to have the general form \cite{AFS,BK}
\be
\sigma_{12} =\exp i \big[ \theta_{12} \big]
= \exp \Big[ i \, \sum_{r=2}^\infty
\sum_{s=r+1}^\infty
c_{r,s} \left( q_r(\xpm{1})\,q_s(\xpm{2})-q_s(\xpm{1})\,q_r(\xpm{2}) \right) \Big] \ ,
\ee
with $q_r(x)$ the conserved magnon charges, defined through
\be
q_r(\xpm{})=\frac{i}{r-1} \, \left( \frac{1}{(\xp{})^{r-1}}-\frac{1}{(\xm{})^{r-1}} \right) \ ,
\ee
and $c_{r,s}$ some coefficients depending on the coupling constant $g = \sqrt{\l}/4\pi$, with 
$\l$ the 't~Hooft coupling. A strong-coupling expansion for the phase $\theta_{12}$ has been 
proposed in~\cite{BHL}, 
\be
c_{r,s} = \sum_{n=0}^{\infty} c_{r,s}^{(n)} g^{1-n} \ ,
\label{strongc}
\ee
with the coefficients given by 
\be
c_{r,s}^{(n)} \, = \, (r-1)(s-1) \bernoulli_{n} \, {\cal A}(r,s,n) \ ,
\ee
where $\hbox{B}_n$ denotes the $n$-th Bernoulli number, and 
\be
{\cal A}(r,s,n) = \frac{\big((-1)^{r+s}-1\big)}
{4\cos(\half\pi n)\,\Gamma[n+1]\,\Gamma[n-1]}\,
\frac{\Gamma[\half(s+r+n-3)]}{\Gamma[\half(s+r-n+1)]}\,
\frac{\Gamma[\half(s-r+n-1)]}{\Gamma[\half(s-r-n+3)]} \ ,
\ee
which vanishes when $r+s$ is even or if $n \geq s-r+3$. This expression agrees with the 
perturbative expansion for strings in $AdS_5 \times S^5$ at leading order \cite{AFS}, 
and includes the first quantum correction \cite{HL}. 
Recently an educated guess was suggested in \cite{BES} for the weak-coupling expansion 
coefficients,
\be
c_{r,s} = \sum_{n=0}^{\infty} \tilde{c}^{(n)}_{r,s} g^{n+1} \ ,
\label{weak}
\ee
that leads to a violation of BMN-scaling at four-loop order, in remarkable agreement with 
the results of \cite{Bern}. Moreover, the conjecture in \cite{BES} still preserves 
the Kotikov-Lipatov transcendentality principle. The aim of this note we will be to derive 
the weak-coupling coefficients in (\ref{weak}) and the pattern of transcendentality 
by a {\em non-perturbative prescription} for resummation of the asymptotic series 
defining the dressing phase factor in the strong coupling regime. The non-perturbative 
prescription reproducing the result in \cite{BES} and~\cite{Bern} is formally the same 
used to define non-perturbatively the $c=1$ matrix model \cite{Gross}. Moreover, the 
dressing phase factors at leading order can be interpreted in terms of a modified $c=1$ 
matrix model.
  
\no
{\bf Weak-coupling expansion.} Let us start writing a convenient symmetrization for the 
strong-coupling expansion of the dressing phase factor \cite{AF}, 
\ba
\theta_{12} = 
& \!\!\! + \!\!\! & \chi(\xp{1},\xp{2}) - \chi(\xp{1},\xm{2}) - \chi(\xm{1},\xp{2}) + \chi(\xm{1},\xm{2})
\nonumber \\
& \!\!\! - \!\!\! & \chi(\xp{2},\xp{1}) + \chi(\xm{2},\xp{1}) + \chi(\xp{2},\xm{1}) - \chi(\xm{2},\xm{1}) \ .
\ea
where
\be
\chi(x_1,x_2)= - \sum_{r=2}^\infty\sum_{s=r+1}^\infty
\frac{c_{r,s}}{(r-1)(s-1)}\,\frac{1}{x_1^{r-1}x_2^{s-1}} \ .
\ee
At strong-coupling we get, from (\ref{strongc}),
\be
\chi(x_1,x_2) = \sum_{n=0}^{\infty} \chi^{(n)}(x_1,x_2) \, \frac {\bernoulli_{n}}{g^{n-1}} \ , 
\label{strong}
\ee
with 
\be
\chi^{(n)} (x_1,x_2) = - \sum_{r=2}^{\infty} \sum_{s=r+1}^{\infty} 
\frac {{\cal A}(r,s,n)}{x_1^{r-1}x_2^{s-1}} \ .
\ee
The strong-coupling expansion (\ref{strong}) is an asymptotic expansion. As it contains the 
Bernoulli numbers $\hbox{B}_n$, which grow like $n !$, it is highly divergent. However, it can still 
be defined non-perturbatively in a similar way to the one used in \cite{Gross} for the non-perturbative 
definition of the $c=1$ matrix model. In order to show this, we will first introduce some new variables
\be
\mu_i \equiv x_i g \ .
\ee
In terms of these variables
\be
\chi(\mu_1,\mu_2) = \sum_{\alpha} g^{\a} \left( \sum_{r,s} 
\frac {\hbox{B}_{r+s-1-\alpha} \, {\cal A}(r,s,r+s-1-\alpha)}{\mu_1^{r-1}\mu_2^{s-1}} \right) \ .
\label{chi}
\ee
The leading order term in (\ref{chi}) is
\be
\chi(\mu_1,\mu_2)^{\hbox{\tiny{LO}}} = g^2 \sum_s \frac {\hbox{B}_{s-1} {\cal A}(2,s,s-1)}{\mu_1\mu_2^{s-1}}
\equiv \frac {g^2}{2\mu_1} \left( \sum_{n=2} \frac { i^n \hbox{B}_n }{n \mu_2^n} \right) \ .
\label{sum}
\ee
The procedure we will now apply is as follows: We first will try to evaluate the sum in 
(\ref{sum}) performing a Borel transform. However the Borel transform contains an infinite 
number of poles on the real axis and the series is thus non-summable, unless a non-perturbative 
prescription is chosen in order to evaluate the integral. This prescription introduces an 
infinite number of parameters. Following a principal value prescription as in the $c=1$ 
matrix model, the non-perturbative definition of (\ref{sum}) will provide 
a perfectly convergent weak-coupling expansion in powers on $\mu_2$ of the form 
$\sum \tilde{c}_n \mu_2^n$, for some coefficients $\tilde{c}_n$. The final step will be the 
derivation of this weak-coupling expansion from the general expression for $\chi(x_1,x_2)$, 
but now using the weak-coupling expansion (\ref{weak}) for the coefficients in the dressing 
phase. This provides an explicit expression for the weak-coupling coefficients $\tilde{c}_{r,s}^{(n)}$ 
in (\ref{weak}) in terms of the coefficients $\tilde{c}_n$ derived from the non-perturbative 
prescription. Let us now perform all these steps.
  
In order to evaluate the sum in (\ref{sum}) we rewrite 
\be
\frac {i^n \hbox{B}_n}{n \mu_2^n} 
= \int_0^{\infty} dt e^{-i \mu_2 t} \frac {\hbox{B}_n}{n !} t^{n-1} \ ,
\ee
so that the Borel transform is 
\ba
\int_0^{\infty} dt e^{-\mu_2 t} \sum_{k=1}^{\infty} (-1)^k \frac {\hbox{B}_{2k}}{(2k)!} t^{2k-1}
= \int_0^{\infty} dt e^{-\mu_2 t} \left( \frac {1}{2} \cot(t/2) - \frac {1}{t} \right) \ .
\ea
The Borel transform does not exist because the integrand has an infinite number of poles 
on the real axis. Therefore in order to find the sum some integration prescription around each 
pole needs to be specified. Such a prescription is interpreted as a non-perturbative 
definition of the sum. In particular, in order to evaluate the integral we will use 
a principal value prescription. From the sum of residues we then get
\be
\pi \sum_{n=1}^{\infty} e^{-2\pi \mu_2 n} = \pi \left( \frac {1}{2\pi\mu_2} - \frac {1}{2} 
+ \sum_{k=1}^{\infty} \frac {\hbox{B}_{2k}}{(2k)!} (2 \pi \mu_2)^{2k} \right) = 
\frac {\pi}{2} \big( \coth (\pi \mu_2) - 1 \big) \ .
\ee
Using now that
\be
\Psi(z) = - \, \gamma - \frac {\pi}{2} \cot (\pi z) \ , 
\ee
we finally get \footnote{The non-perturbative result (\ref{npsum}) slightly differs from 
the resummation in \cite{BHL}, where the integration was performed with a rotation, 
$\int_0^{\infty} dt e^{-i \mu t} (1/2 \coth(t/2)-1/t) = - \gamma + i/2\mu 
+ \log(i\mu) - \Psi(i\mu)$. This difference is crucial in order to recover the 
Yang-Mills phase factor at second order. In fact (\ref{npsum}) is the exact analogue 
of the matrix model solution \cite{Gross}.}
\be
\chi(\mu_1,\mu_2)^{\hbox{\tiny{LO}}} = - \frac {g^2}{2\mu_1} 
\big( \gamma + \Re \, \Psi(\mu_2) + \frac {\pi}{2} \big) \ ,
\label{npsum}
\ee
where $\Re \, \Psi(\mu)$ denotes the real part of $\Psi(\mu)$. From the non-perturbative 
expression (\ref{npsum}) we get the following weak-coupling expansion, convergent for $|\mu_2|<1$,
\ba
\chi(\mu_1,\mu_2)^{\hbox{\tiny{LO}}} \!\!\! & = & \!\!\! 
- \frac {g^2}{2\mu_1} \big( \gamma + \pi/2 \big) - \frac {g^2}{2\mu_1} \big( \gamma - \sum_{k=1} 
(-1)^k \zeta(1+2k) \mu_2^{2k} \big) \nonumber \\
\!\!\! & = & \!\!\! \frac {g^2}{2\mu_1} \sum_{k=1}^{\infty} (-1)^k \zeta(1+2k) \mu_2^{2k} 
- \frac {g^2}{2\mu_1} \frac {\pi}{2} \ .
\label{zeta}
\ea
In order to get the weak-coupling coefficients in the dressing factor we will write 
$\chi(x_1,x_2)$ in the weak-coupling regime (\ref{weak}) for the unknown coefficients 
$\tilde{c}^{(n)}_{r,s}$,
\be
\chi(x_1,x_2) = \sum_{r,s,n} g^{n+1} \frac {\tilde{c}^{(n)}_{r,s}}{(r-1)(s-1)} 
\frac {1}{x_1^{r-1}x_2^{s-1}} \ .
\ee
Comparison with (\ref{zeta}) requires considering the $r=2$, $n=s-1$ piece, which leads to 
\be
\frac {g}{x_1} \sum_s \tilde{c}_{2,s}^{(s-1)} \frac {g^{s-1}}{x_2^{s-1}} = 
\frac {g}{x_1} \sum_s \frac {\tilde{c}_{2,s}^{(s-1)}}{(s-1)} (gy_2)^{s-1} \ ,
\ee
that we can compare with (\ref{zeta}) for $gy_2=\mu_2$. From this we get the result 
\be
\tilde{c}_{2,s}^{(s-1)} = (-1)^{(s-1)/2} \zeta(s) \frac {(s-1)}{2} 
\label{tildec}
\ee
for $s$ odd, and 
\be
\tilde{c}_{2,s}^{(s-1)} = 0 
\ee
for $s$ even. This result is in complete agreement with the Kotikov-Lipatov transcendentality 
principle.
  
\no
{\bf Non-perturbative prescription.} Let us now briefly elaborate on the non-perturbative 
ambiguity. As we have already discussed the Borel transform does not properly exist due to 
the infinite number of poles along the integration range. The non-perturbative prescription 
that we have employed above is based on the Cauchy principal part. However, we must recall 
that the general procedure in order to give meaning to an infinite integral, in the distribution 
sense, is to first define a regularized distribution on the space of test functions, 
with support away from the singularities. Then an extension to the whole space of test functions 
is constructed. This extension in general does exist, but it is not unique. In our case a 
simple way to parameterize the intrinsic ambiguity in the definition of the infinite integral 
as a distribution is including a distribution of the type 
\be
2 \pi \sum_i c_i \, \delta(x-x_i) \ ,
\ee
with $x_i$ the location of the (simple) poles. The coefficients $c_i$ are thus the non-perturbative 
parameters that we should fix from some alternative non-perturbative definition of the theory. 
In the absence of such an alternative definition we are unfortunately forced to deal with all 
these free constants. An economic possibility is to have all the $c_i$ equal to some arbitrary 
constant $\a$, with $\a=1$ corresponding to the Cauchy principal part. This implies a 
modification of (\ref{tildec}) to
\be
\tilde{c}_{2,s}^{(s-1)} = (-1)^{(s-1)/2} \, \a \, \zeta(s) \frac {(s-1)}{2} \ ,
\ee
and it therefore looks that at least to fourth order $\a=4$ is the right non-perturbative 
prescription \cite{Bern}. However, we should still keep in mind that any violation of this guess 
at higher orders will only force a different choice of the arbitrary parameters $c_i$.
  
\no  
{\bf Discussion.} One interesting aspect of the result (\ref{npsum}) for the phase factor is 
the very strong analogy with the $c=1$ matrix model. Using the relation of the digamma function 
and Hurwitz zeta function,
\be
\lim_{s \rightarrow 1} \left[ \zeta(s,z) - \frac {1}{s-1} \right] = - \Psi(z) \ ,
\ee
with 
\be
\zeta(s,z) = \sum_{n=0}^{\infty} \frac {1}{(n+z)^s} \ ,
\ee
we can map the dressing phase factor (\ref{npsum}) with the density of states $\rho(\mu)$ of a 
matrix model, with the only {\em important} difference that instead of using the harmonic oscillator 
energy spectrum, $(n+1/2)\omega \hbar$, we now have $n\omega \hbar$. In matrix models the 
density $\rho(\mu)$ is related to the phase shift introduced in the wave function by the matrix 
potential through $\rho(\mu)=\partial \delta(\mu)/\partial \mu$. In this sense it looks like some 
parts of the dressing phase factor entering the integrable spin chain description of planar 
${\cal N}=4$ supersymmetric Yang-Mills could be related to a phase shift in some matrix model 
through a formal relation of the form $\delta_{{\cal N}=4}=\partial \delta_{\hbox{\tiny{matrix}}}$. 
  
The analogy with the matrix model goes a bit further if we consider the strong and weak-coupling 
expansions of the density $\rho(\mu)$. In fact both map, respectively, into the weak 
and strong-coupling expansions of the dressing phase factor. Moreover, as it seems to be the 
case for the dressing phase factor, the $c=1$ weak-coupling expansion is only asymptotic, 
while the strong-coupling one is perfectly convergent. This potential connection with $c=1$ 
matrix models certainly deserves future research.~\footnote{An appealing probe for the matrix 
analogy could be the relation of a finite-temperature matrix model to the finite-size 
exponential corrections in the quantum string energy shift \cite{SZZ}.}


\vspace{4mm}
\centerline{\bf Acknowledgments}

We are grateful to A.~Delgado and A.~Sabio Vera for discussions. The work of C.~G. is partially 
supported by the Spanish DGI contract FPA2003-02877 and CAM project HEPHACOS P-ESP-00346. 


\end{document}